\documentstyle[prl,aps,twocolumn,epsf]{revtex}
\newcommand{\lapx}{\mbox{\raisebox{-4pt}{$\,\buildrel<\over\sim\,$}}}
\newcommand{\gapx}{\mbox{\raisebox{-4pt}{$\,\buildrel>\over\sim\,$}}}
\begin{document}
\draft

\title{Crossover from nonadiabatic to adiabatic electron transfer
reactions: Multilevel blocking Monte Carlo simulations}
\author{L.~M{\"u}hlbacher$^1$ and R.~Egger$^2$}
\address{${}^1$Fakult{\"a}t f{\"u}r Physik,
   Albert-Ludwigs-Universit{\"a}t,
   D-79104 Freiburg, Germany\\
  ${}^2$Institut f{\"u}r Theoretische Physik,
   Heinrich-Heine-Universit{\"a}t,
   D-40225 D{\"u}sseldorf, Germany}
\date{Date: \today}

\maketitle

\begin{abstract}
The crossover from nonadiabatic to adiabatic electron transfer has been
theoretically studied under a spin-boson model (dissipative two-state system)
description. We present numerically exact data for the thermal transfer rate
and the time-dependent occupation probabilities in largely unexplored regions
of parameter space, using real-time path-integral Monte Carlo simulations. The
dynamical sign problem is relieved by employing a variant of the recently
proposed multilevel blocking algorithm. We identify the crossover regime
between nonadiabatic and adiabatic electron transfer, both in the classical
(high-temperature) and the quantum (low-temperature) limit. The electron
transfer dynamics displays rich behaviors, including multi-exponential decay
and the breakdown of a rate description due to vibrational coherence.
\end{abstract}
%\pacs{}

\newpage

\narrowtext

\section{Introduction}

Electron transfer (ET) reactions are important for an understanding of many
processes in chemistry, biology and physics. Popular examples include charge
transfer in semiconductors, chemical reactions, or the primary ET step in
bacterial photosynthesis, and, despite the long history of the field
\cite{marcus56,marcus85}, ET still attracts unbroken and (probably
exponentially) increasing attention \cite{kuznetsov,tributsch}. In this paper,
a theoretical study of ET reactions based on the spin-boson (dissipative
two-state) description is presented. The spin-boson model provides a
well-established description of ET processes in condensed phase environments,
which can be motivated from the central limit theorem or, alternatively, using
linear response theory \cite{chandler,song,weiss}. In this model, two localized
quantum states corresponding to the redox sites are taken. This two-level
system (TLS) is fully specified by a tunnel splitting $\hbar\Delta$ (which is
twice the electronic coupling between the two redox sites), and by a bias
$\hbar\epsilon$ (which corresponds to the energy difference between the
sites). Of crucial importance is then the coupling to the ``environmental''
modes (bath), which is modeled as an infinite set of harmonic oscillators with
a suitable (continuous) spectral density $J(\omega)$ describing their
frequency-resolved coupling strength to the electron. For a given ET system,
one can use classical Molecular Dynamics simulations to find the appropriate
$J(\omega)$ \cite{chandler}, but below we shall consider a specific class of
spectral densities that describes many ET processes quite
accurately. Two important
parameters following from a given spectral density are the {\sl reorganization
energy} $\hbar\Lambda$ (which describes the overall coupling strength) and the
typical bath frequency $\omega_c$ corresponding to a maximum in
$J(\omega)$. This model has been studied in depth in the field of dissipative
quantum mechanics \cite{weiss,feynman,grabert,VCM,hanggi}, mostly in the
so-called scaling limit for Ohmic spectral densities (see below), but its
application to ET reactions is mainly located in a different parameter regime.

Certain limiting cases of ET theory are rather well understood. The primary
object of interest is the {\sl thermal rate constant} $k_{th}$, which is useful
if ET dynamics is correctly described by exponential decay. First, for very
small $\Delta$, one can use perturbation theory in $\Delta$ to obtain the
celebrated golden rule rate \cite{levich,ovch,voro70,GR2}. This is the {\sl
nonadiabatic limit}. In the opposite limit of large $\Delta$, the {\sl
adiabatic limit}, it is more useful to think in terms of adiabatic free energy
surfaces formed by linear superpositions of the two localized free energies. In
that case, the bath essentially is very slow (``classical'') compared to the
electronic motion, and one can develop a simple theoretical description again
\cite{carmeli,gehlen,burshtein}. Notably, the ET dynamics in this regime is
often not simply an exponential relaxation but exhibits oscillatory or
multi-exponential behavior at low temperatures \cite{lothar1,hornbach}. At
sufficiently high temperatures, the famous {\sl classical Marcus theory}
\cite{marcus56,marcus85}, or semiclassical generalizations thereof
\cite{kuznetsov}, apply. For a path-integral approach to Marcus theory, see
also Ref.~\cite{garg85}. Marcus theory also makes predictions about the rate
constant in the crossover regime between nonadiabatic and adiabatic ET. It
should be stressed that the spin-boson model contains all these results as
limiting cases in the full parameter space.

Notably, the spin-boson problem cannot be solved analytically without imposing
approximations. One important exception to this rule is the case of an Ohmic
spectral density, see Eq.~(\ref{ohmic}) below, with the restrictions:
\begin{equation}\label{scallimit}
 \Delta/\omega_c\ll 1 \, , \quad k_B T/\hbar\omega_c \ll 1
\quad \Leftrightarrow \quad \mbox{scaling limit} \;.
\end{equation}
In this scaling limit, sophisticated conformal field theory techniques are able
to yield the complete non-equilibrium solution for the dynamics under arbitrary
initial preparation \cite{saleur}. Besides this solution, whose parameter regime
is of little interest to ET theory, if one is interested in analytical
progress, one has several options for approximations. One possibility is to
work perturbatively in the tunnel splitting $\Delta$. This leads to the
noninteracting blip approximation (NIBA) \cite{leggett,nica,milena}, which is
basically equivalent to the golden rule formula for the rate constant. Other
approaches are based on variational treatments \cite{silbey} or work
perturbatively in the system-bath coupling (Redfield approach) \cite{friesner}.
As typical reorganization energies are quite sizeable, a perhaps more promising
route is to use the powerful computer simulation methods available by now.

Let us then briefly summarize the numerical methods that have been applied to a
study of the spin-boson dynamics. An approximate approach which has become
quite popular recently employs a mixed quantum/classical (or
quantum/semiclassical) description
\cite{stock1,stock2,stock3,sun,miller,lucke,golosov1,golosov2,casado,thoss}.
However, such studies are typically restricted to a few bath modes. Unless one
has a situation with only a few strongly coupled bath modes, the more typical
situation of condensed-phase ET reactions is difficult to describe using
such methods. In addition, it is hard to estimate the errors made under these
approximations. Similar arguments apply to time-dependent Hartree methods
(basis set approach) \cite{wang} and to memory-equation approaches
\cite{nancy,winterstetter}. The latter method, as well as the recently proposed
stochastic Langevin-type simulation \cite{stockburger1,stockburger2}, dynamical
Wilson renormalization group \cite{costi1,costi2} and the real-time
renormalization group \cite{schoeller}, basically appear to be restricted to the
Markovian limit realized for an Ohmic bath with $\omega_c\to \infty$, yet most
situations of interest for ET are located outside the Markovian regime. The
methods discussed so far are therefore very powerful in certain parameter
regions of the spin-boson model (in fact much more powerful than our method
employed below), but may fail badly in others.

A method free from such restrictions is provided by {\sl Monte Carlo
simulations} \cite{qmcgeneral}, in particular the powerful path-integral Monte
Carlo (PIMC) approach. In principle, this approach is numerically exact
throughout the full parameter space, i.e.~for arbitrary spectral density,
tunnel coupling, temperature and/or bias. However, as one is really interested
in the ET {\sl dynamics} in real time, a direct application of quantum Monte
Carlo (QMC) methods is plagued by the infamous {\sl sign problem} \cite{loh}.
One alternative is to carry out an imaginary-time simulation, which is
intrinsically free of the sign problem, and then analytically continue the data
to real time. This program has been carried out for the spin-boson problem,
either using Pade approximants \cite{sudip,voelker} or Maximum Entropy methods
\cite{bailey} in order to perform the analytical continuation. The latter
procedure is mathematically ill-defined and quite troublesome to carry out. In
addition, by using this procedure, one is inevitably restricted to equilibrium
situations. However, in ET reactions one often has a true non-equilibrium
preparation, e.g.~one photo-excites an electron to the donor state at time
$t=0$ and then follows the time evolution of the occupation probability. Such a
question has to be answered by a true non-equilibrium real-time calculation.

Such a calculation is presented in our study. The price to pay for working in
real time and still keeping exactness and applicability throughout the full
parameter space is however quite high. The main challenge is due to the sign
problem, which implies that one is restricted in the real-time range up to
which stable calculations can be done. The sign problem arises from the
destructive interference of a large number of quantum paths at long times,
leading to a very small signal-to-noise ratio and thus rendering the
simulations numerically unstable. Early real-time PIMC attempts for computing
the spin-boson dynamics \cite{behrman,mak1,egger94} were based on the
stationary-phase Monte Carlo approach \cite{filter,spmc,acp} or variations
thereof. Such filtering methods try to keep the MC trajectory close to the
important stationary-phase regions of path space. While this is
able to somewhat relieve the sign problem, a more powerful
method is available by now. This ``multilevel blocking'' (MLB) algorithm
\cite{mlb98,mlb99,mlb00,mischa} has been proposed and successfully used to
relieve both the fermionic and the dynamical sign problem. MLB represents a
systematic recursive implementation of a simple {\sl blocking
strategy}. Roughly speaking, the blocking strategy is equivalent to a naive
filtering, and one can think of MLB as a systematic and optimized way of
implementing the filtering idea. Here we shall employ a recently proposed MLB
variant which can account for effective actions \cite{mlb00} of the type
encountered in PIMC for the spin-boson model. This approach is superior to
previous real-time PIMC methods, and allows to compute the spin-boson dynamics
up to timescales of practical interest {\sl without approximations and for the
full parameter regime relevant to ET reactions}. 

The remainder of this paper is as follows. In Sec.~\ref{theory} we briefly
describe the spin-boson model and the dynamical quantities of interest. Next we
describe our QMC method, see Sec.~\ref{MC}. Since the method has been exposed
in great detail in Ref.~\cite{mlb00}, focusing on exactly the same model, we
largely refrain from repeating ourselves here. 
Technical details of interest to experts are given in the Appendix.
In Section \ref{results} we then present our numerical
results for the dynamical quantities and the rate $k_{th}$ for various regimes,
including the important crossover region between nonadiabatic and adiabatic
ET. The most difficult regime is at low temperatures, a regime that can now be
treated by PIMC. In presenting the numerical results, we restrict ourselves to
the symmetric (unbiased) case $\epsilon=0$, leaving the biased system for
future studies. Finally, some conclusions are drawn in Sec. \ref{conclusion}.

\section{Electron transfer theory} \label{theory}

The spin-boson model (dissipative two-state system) is defined by the
Hamiltonian \cite{weiss}
\begin{eqnarray} \label{hamilton}
H &=& H_0 + H_I + H_B = -{\hbar\Delta\over2}\sigma_x
 + {\hbar\epsilon\over2}\sigma_z \nonumber\\
&&- \sigma_z \sum_\alpha C_\alpha X_\alpha
 + \sum_\alpha \left( {P_\alpha^2 \over 2m_\alpha}
 + {1\over2}m_\alpha\omega_\alpha^2 X_\alpha^2 \right) \;,
\end{eqnarray}
where the free TLS is described by $H_0$, containing the tunnel splitting
$\hbar\Delta$ and the energy gap $\hbar\epsilon$ between the two localized
electronic states. Here $\sigma_x$ and $\sigma_z$ denote the standard Pauli
matrices acting in TLS space, where the $|+\rangle$ ($|-\rangle$) state refers
to the donor (acceptor). The environmental modes are modeled by an infinite
collection of harmonic oscillators, $H_B$, which bilinearly couple to the
position of the electron ($H_I$). For the great majority of ET systems, this
model provides a reasonably accurate description of reality, as is discussed
and motivated at length in Refs.~\cite{chandler,song,weiss,GR2}. Within this
model, the bath parameters affect ET only via the spectral density
\begin{equation}
J(\omega) = \frac{2\pi}{\hbar} \sum_\alpha \frac{C_\alpha^2}
	 {m_\alpha\omega_\alpha} \delta(\omega-\omega_\alpha) \,,
\end{equation}
which effectively becomes a continuous function of $\omega$ for a
condensed-phase environment and determines all bath correlation functions that
are relevant for the ET dynamics. For instance, the complex-valued bath
autocorrelation function for complex time $z=t-i\tau$ is for $\beta=1/k_B T$
given by
\begin{equation} \label{lz}
L(z) = \int_0^\infty \frac{d\omega}{\pi} J(\omega)
{\cosh[\omega(\hbar\beta/2-iz)] \over \sinh(\hbar\beta\omega/2)} \;.
\end{equation}
Similarly, the important {\sl reorganization energy} $\hbar \Lambda$ of Marcus
theory \cite{marcus56,marcus85} is defined by
\begin{equation} \label{reorg}
\Lambda = \int_0^\infty d\omega\ {J(\omega) \over \pi\omega} \, ,
\end{equation}
which is the only important bath quantity in the classical (high-temperature)
limit, as is apparent from classical Marcus theory. Details of $J(\omega)$
matter only at low temperatures.

In the following we focus on the two dynamical properties of primary relevance
to ET dynamics. The {\sl time-dependent occupation probability} is defined as
\begin{equation} \label{P(t)}
P(t) = \langle\sigma_z(t)\rangle = \langle e^{iHt/\hbar} \sigma_z
e^{-iHt/\hbar} \rangle
\end{equation}
and gives the difference in the occupation probabilities of the donor and the
acceptor state with the electron initially held fixed on the donor. This
quantity then directly probes ET dynamics after the non-equilibrium initial
preparation corresponding to the initial density matrix
\begin{equation} \label{initial_perp_P}
W(0) =  |+\rangle\!\langle+| e^{-\beta(H_B+\mu{\cal E})}\;.
\end{equation}
Here $\mu$ is the dipole moment of the electron, and ${\cal E}$ denotes the dynamical
polarization of the bath \cite{weiss}, with the electron in the donor state and
the bath equilibrated with respect to it.  By comparing with
Eq.~(\ref{hamilton}), we see that $\mu {\cal E}=-\sum_\alpha C_\alpha
X_\alpha$.  As pointed out in Ref.~\cite{lucke}, this ``standard preparation''
often used in ET experiments is unfavorable for coherent (oscillatory) dynamics
when compared to other initial preparations. Therefore it is especially
suitable for a thermal transfer rate description.

In addition, we calculate the complex-valued {\sl correlation function}
\begin{eqnarray} \label{corr}
C(t) &=& \langle\sigma_z(0)\sigma_z(t)\rangle_\beta \nonumber\\
&=& Z^{-1}{\rm tr}\left\{
 e^{-\beta H}\sigma_z e^{iHt/\hbar}\sigma_z e^{-iHt/\hbar}
 \right\}
\end{eqnarray}
with $Z = {\rm tr} \, \exp(-\beta H)$. This quantity probes the dynamics under
an equilibrium preparation and, for $\epsilon=0$, differs from $P(t)$ only through the initial
preparation. At high temperatures, one generally expects that initial
preparation effects show little influence on the dynamics, and therefore a
unique rate constant $k_{th}$ should exist. The inverse rate would then be the
timescale governing the relaxation of both $P(t)$ and $C(t)$. However, in the
quantum regime one needs to be more careful, and the specific experiment under
study will determine which is the relevant dynamical quantity.

Assuming for the moment that there is a unique rate constant $k_{th}$, $P(t)$
should follow a simple exponential decay for times $t$ large compared to some
transient molecular timescale $\tau_{\rm trans}$ of the order $1/\omega_c$, yet
small compared to the relaxation timescale $\tau_{\rm relax}\approx
1/k_{th}$. Calculating $P(t)$ for $\tau_{\rm trans} \ll t \ll \tau_{\rm relax}$
would then allow to extract the temperature-dependent thermal transfer
rate. The thermal transfer rate (if it exists) can also be obtained via a {\sl
time-dependent function} defined under an equilibrium preparation \cite{VCM},
\begin{equation} \label{timerate}
k_f(t) = {2 \over \hbar\beta Z_A}{\rm Im} \, {\rm tr}
\left[ e^{-\beta H} h_A(0) h_A(t) \right] \,,
\end{equation}
where $h_A = |+\rangle\!\langle+|= (1+\sigma_z)/2$ denotes a projector onto the
donor state and $Z_A = {\rm tr}\{e^{-\beta H}h_A\}$ its partition sum. The
thermal rate constant then follows from Eq.~(\ref{timerate}) as the {\sl
plateau value} for times $t$ where $k_f(t)$ forms a plateau. According to our
above discussion, this should happen for $\tau_{\rm trans} \ll t_{\rm pl} \ll
\tau_{\rm relax}$ where $t=t_{\rm pl}$ is a suitable plateau time. The index
$f$ in Eq.~(\ref{timerate}) denotes a {\sl forward rate}, describing the
directed transfer rate from the donor to the acceptor site. The total rate
$k_{th}$ probed by $P(t)$ is the sum of the forward and the backward rate $k_f$
and $k_b$, which in turn are connected via a standard detailed-balance
relation,
\begin{equation}
k_b(T,\epsilon)= k_f (T,\epsilon) \exp(-\hbar\epsilon/k_B T) \,.
\end{equation}
Notably, if $\tau_{\rm trans}$ and $\tau_{\rm relax}$ are not well separated,
no  transfer rate can be defined, and a simple rate formalism
breaks down. Such a breakdown of the rate description could happen, for
instance, for very fast reactions. Equation (\ref{timerate}) thus not only
yields a convenient way to compute the thermal transfer rate, but also a means
to decide upon the validity of the rate picture of ET at all. It is important
to stress that $k_f(t)$ does not refer to a time-dependent rate reflecting the
dynamics of $P(t)$ but is only an auxiliary function allowing to determine the 
time-independent rate $k_{th}$ when it exists.  Equation
(\ref{timerate}) can be expressed in terms of $C(t)$,
\begin{equation} \label{timerate2}
k_f(t) = {1 \over \hbar\beta}
\frac{{\rm Im}\ C(t)}{1+\langle\sigma_z\rangle_\beta} \, ,
\end{equation}
with $\langle\sigma_z\rangle_\beta = Z^{-1}{\rm tr}\left\{e^{-\beta
H}\sigma_z\right\}$. The fact that $P(t)$ will follow an exponential decay only
after some time $\tau_{\rm trans}$ is made explicit by noting that $dP/dt (t=0)
= 0$ due to $d\sigma_z(t)/dt \sim \sigma_y(t)$. Our main goal is therefore to
calculate the ET dynamics up to real times that are significantly larger than
$\tau_{\rm trans}$.

Before turning to the numerical computations, let us briefly summarize some
simple limiting cases, where analytical results are available. In the {\sl
nonadiabatic limit}, Eq.~(\ref{timerate}) can be evaluated for arbitrary
temperatures using Fermi's golden rule,
\begin{equation} \label{k3}
k^{GR}_f = {\Delta^2\over4}\int_{-\infty}^\infty dt\,
\exp[i\epsilon t -Q(t)]\,,
\end{equation}
with the twice-integrated bath autocorrelation function:
\begin{equation}\label{qz}
d^2Q(z)/dz^2 = L(z) \qquad \mbox{with} \quad Q(0) = 0 \,.
\end{equation}
For special choices of $J(\omega)$, it is possible to evaluate the remaining
integral explicitly \cite{GR2}. In the {\sl adiabatic limit}, the bath is very
slow and thus behaves classically. Essentially, it then represents a static
random field obeying Gaussian statistics, which acts as an additional
fluctuating energy gap between the electronic states. The ET dynamics then
follows from the free TLS dynamics by including a fluctuating bias
\cite{chandler}.  The inevitable failure of this approach at long times
can in principle be relieved by applying appropriate Bloch-type equations
\cite{carmeli}. Third, for high
temperatures, {\sl classical Marcus theory} \cite{marcus85} can be recovered
from the spin-boson description \cite{garg85} and predicts the forward rate
\begin{equation} \label{marcus}
k^{cl}_f = \frac{\Delta^2}{4+\pi\Delta^2/(\Lambda\omega_r)}
 \sqrt{{\pi \hbar \over \Lambda k_B T}}\, e^{-F^\ast(\epsilon)/k_B T} \,,
\end{equation}
with the activation free energy barrier given by the celebrated {\sl Marcus
parabola},
\begin{equation} \label{Fstar}
F^\ast(\epsilon) = \hbar(\epsilon - \Lambda)^2/4\Lambda \,,
\end{equation}
and a solvent frequency scale $\omega_r$.  For the Ohmic spectral density
(\ref{ohmic}), this frequency was computed approximately in Ref.~\cite{garg85},
with the result
\begin{equation} \label{omegar}
\omega_r \approx \frac{\omega_c}{2} \,.
\end{equation}
Marcus theory yields a classical rate constant covering the full crossover from
nonadiabatic to adiabatic ET. Equation (\ref{marcus}) coincides with the
high-temperature limits of the nonadiabatic rate (\ref{k3}) and of the
adiabatic formulation for $\Delta\ll \omega_c$ and $\Delta\gg \omega_c$,
respectively. In the adiabatic limit, the rate is independent of $\Delta$.
>From Eq.~(\ref{marcus}), the crossover regime is expected for $\Delta/\omega_c
\approx \sqrt{\Lambda/\omega_c}$, at least for high temperatures.

In all simulations reported in this paper, we consider an {\sl Ohmic spectral
 density} with exponential cutoff \cite{weiss},
\begin{equation} \label{ohmic}
J(\omega) = 2\pi\alpha\omega\, e^{-\omega/\omega_c} \;,
\end{equation}
with dimensionless damping strength $\alpha$ and a cutoff frequency $\omega_c$.
For many polar solvents, it is appropriate to choose $\omega_c$ according to
some intermediate bath frequency \cite{GR2}. Instead of $\alpha$, it is often
more convenient to measure the coupling strength to the bath in terms of the
reorganization energy (\ref{reorg}). For this spectral density, one finds
$\Lambda = 2\alpha\omega_c$. In addition, the correlation function (\ref{qz})
can be given in closed form \cite{GR2},
\begin{eqnarray}
Q(z) &=& 2\alpha\Bigg[ \ln(1+i\omega_c z) \nonumber\\ \label{qz1}
&& - \ln\left( \frac{\Gamma(\Omega+iz/\hbar\beta)\Gamma(\Omega-iz/\hbar\beta)}
		 {\Gamma^2(\Omega)} \right) \Bigg] \;,
\end{eqnarray}
with the gamma function $\Gamma(z)$ and the abbreviation
\begin{equation}\label{omega}
\Omega=1+k_B T/\hbar\omega_c \, .
\end{equation}
Note that the second contribution in $Q(z)$ in Eq.~(\ref{qz1}) vanishes for
$T=0$.

Let us also summarize at this point some of the analytically known results for
this class of spectral densities. First, the complete temperature dependence of
the nonadiabatic golden rule rate (\ref{k3}) is known for $\alpha=1/2$ and
$\alpha=1$. For $\epsilon=0$, the result is \cite{GR2}
\begin{eqnarray} \label{a12}
k^{GR}_f (\alpha=1/2) &=& {\pi\Delta^2 \over \omega_c}
\frac{\Gamma(2\Omega-1)}{2^{2\Omega}\Gamma^2(\Omega)} \;,\\
\label{a1}
k^{GR}_f (\alpha=1) & =&
 {\pi\Delta^2 \over 2\hbar\beta\omega_c^2}
\frac{\Gamma^4(2\Omega-1)}{\Gamma^4(\Omega)\Gamma(4\Omega-2)} \;.
\end{eqnarray}
Second, the complete dynamics for $\alpha=1/2$ in the scaling limit defined by
Eq.~(\ref{scallimit}) is known. For instance, the quantity $P(t)$ obeys a
simple exponential relaxation for $\epsilon=0$ and $\omega_c t \gg 1$,
\begin{equation} \label{scalingP}
P(t) = \exp\left(-{\pi\Delta^2 \over 2\omega_c}t\right) \;,
\end{equation}
where the relaxation rate is twice the low-temperature limit of
Eq.~(\ref{a12}). Notably, for $\alpha\neq 1/2$ but close to 1/2, the dynamics
is much more complicated, with algebraic decay and oscillatory behaviors
\cite{weiss}. These features are due to {\sl electronic coherence} effects,
which one generally does not expect to survive in ET reactions. This is also
reflected in the behavior of $C(t)$ for $\alpha=1/2$, which exhibits long-time
algebraic tails. Therefore one cannot expect that the time-dependent function
$k_f(t)$ will exhibit plateau behavior, although it is possible to define a
rate for the decay of $P(t)$. This reflects the importance of the initial
preparation in that case.

\section{Simulation method} \label{MC}

\begin{figure}
\epsfxsize=6cm
\epsffile{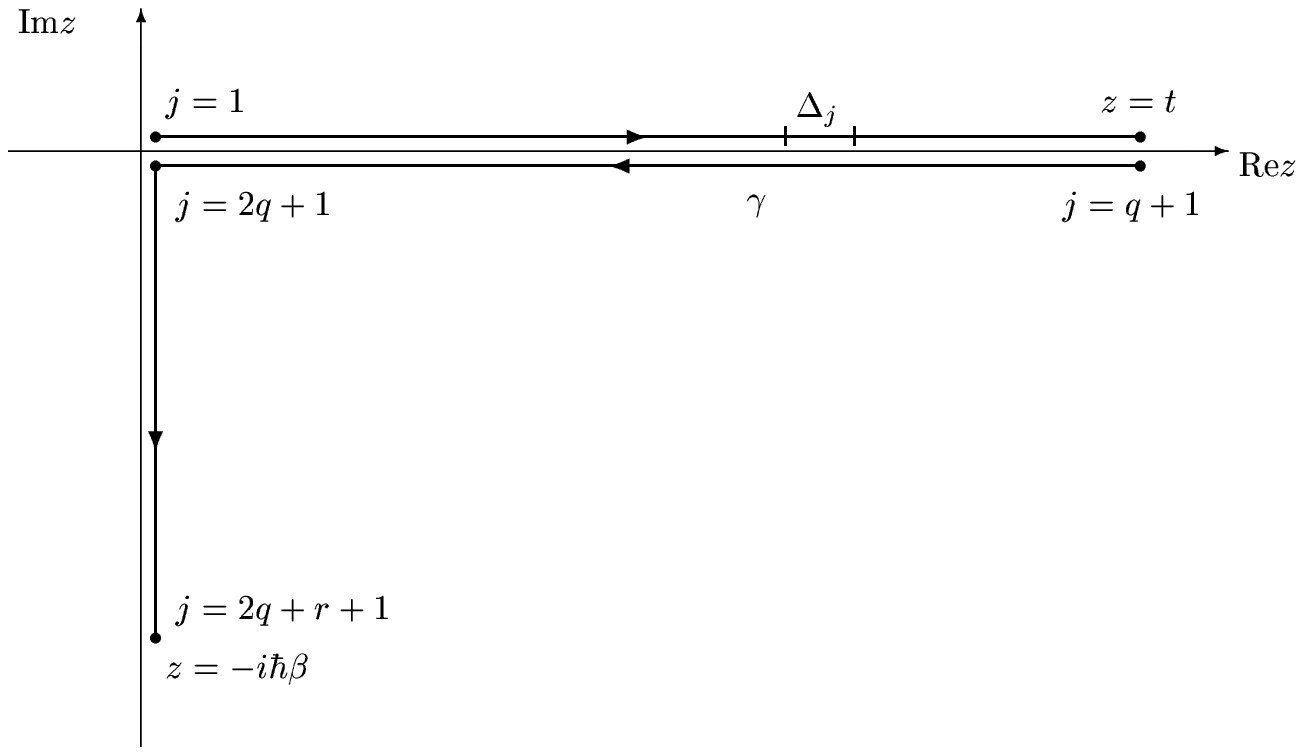}
\caption[]{\label{fig0} Kadanoff-Baym contour $\gamma$ in the complex-time
plane and a possible discretization. }
\end{figure}

Next we turn to a computational method able to yield the rate constant. A great
simplification arises due to the harmonic nature of the bath and the bilinear
coupling in Eq.~(\ref{hamilton}), which allows to perform the trace over the
bath degrees of freedom in Eqs.~(\ref{P(t)}) and (\ref{corr}) in an
\emph{exact} manner. Adopting a path-integral approach, one obtains, e.g.~for
$C(t)$ \cite{weiss},
\begin{equation} \label{corr2}
C(t) = Z^{-1} \int\!{\cal D}\sigma\; \sigma(0)\sigma(t)
 \exp\left\{ {i\over\hbar}S_0[\sigma] - \Phi[\sigma] \right\}
\;.
\end{equation}
Here the path integration runs over paths $\sigma(z)$ for the discrete ``spin''
variable $\sigma=\pm 1$ corresponding to $\sigma_z$, with the complex time $z$
following the Kadanoff-Baym contour $\gamma$ depicted in Fig.~\ref{fig0}.
Furthermore, $S_0[\sigma]$ denotes the total action of the free TLS, and the
influence of the traced-out bath is encoded in the {\sl influence functional}\
$\Phi[\sigma]$. In terms of the bath autocorrelation function (\ref{lz}), it
reads \cite{weiss}
\begin{equation} \label{inflfunct}
\Phi[\sigma] = {1\over4} \int_\gamma dz \int_{z'<z} dz' \sigma(z) L(z-z')
\sigma(z') \,,
\end{equation}
where time integrations are ordered along the contour $\gamma$. The influence
functional introduces long-ranged interactions along the real and imaginary
time axes, rendering the evaluation of the remaining path integral in
Eq.~(\ref{corr2}) a difficult task.

Quantum-mechanical expectation values like Eq.~(\ref{corr2}) can be calculated
in a numerically exact way by PIMC simulations. To make Eq.~(\ref{corr2})
accessible to PIMC, the path integral is taken in its discretized form with $N$
discretization steps. Following the scheme described in
Refs.~\cite{egger94,mlb00}, $q$ uniformly spaced points are used for each of
the real-time paths $z:0 \rightarrow t$ and $z:t \rightarrow 0$, and $r$ points
for the imaginary-time path $z:0 \rightarrow -i\hbar\beta$, with discretization
steps
\begin{equation}
\Delta_j = \left\{
\begin{array}{cc}
t/q, & 1\le j\le q\\
-t/q, & q+1\le j\le 2q\\
-i\hbar\beta/r, & 2q+1\le j\le 2q+r \;,
\end{array}\right.
\end{equation}
leading to $N=2q+r$ discrete time points $z_i=\sum_{j=1}^{i-1} \Delta_j$.
Using $\sigma_i = \sigma(z_i)=\pm 1$, Eq.~(\ref{corr2}) can be written as
\begin{eqnarray} \label{corr3}
C(t) &=& Z^{-1} \sum_{\{\sigma_j\}} \sigma_1\sigma_{q+1} \rho[\{\sigma_j\}] \,, \nonumber\\
&&\rho[\{\sigma_j\}] = \left[ \prod_{i=1}^N K(\sigma_i,\sigma_{i+1}) \right]
e^{-\Phi[\{\sigma_j\}]}
\end{eqnarray}
with $Z=\sum_{\{\sigma_j\}} \rho[\{\sigma_j\}]$ and $\sigma_{N+1}=\sigma_1$ due
to the cyclic nature of the trace. The sum runs over all realizations of the
discretized spin path $\{\sigma_j\} = \{\sigma_1,\dots,\sigma_N\}$, and
\begin{equation} \label{freeprop}
K(\sigma_i,\sigma_{i+1}) =
\langle\sigma_{i+1}|\exp(-i\Delta_i H_0/\hbar)|\sigma_i\rangle
\end{equation}
denotes the short-time propagator of the free TLS, which is a known $2\times 2$
matrix. Note that due to $\langle \sigma(0)\sigma(t) \rangle_\beta =\langle
\sigma(t')\sigma(t'+t) \rangle_\beta$, one single MC trajectory can be used to
compute $C(t_1)$ for all times $t_1\leq t$. Furthermore, exploiting $\langle
\sigma(0)\sigma(t_k) \rangle_\beta = \langle \sigma(t_k)\sigma(0)
\rangle_\beta^\ast$ in a similar way improves the MC statistics. A discretized
version of the influence functional is given by
\begin{equation} \label{inflfunctdisc}
\Phi[\{\sigma_j\}] = {1\over8} \sum_{j,k=1}^N \sigma_j L_{jk}
\sigma_k \;,
\end{equation}
with the complex-valued influence matrix \cite{egger94}
\begin{eqnarray} \label{Lmatrix}
L_{jk} &=& L_{kj}= Q(z_j - z_k + (\Delta_j + \Delta_{k-1})/2) \nonumber\\ && +
Q(z_j - z_k + (-\Delta_{j-1} - \Delta_k)/2) \nonumber\\ && - Q(z_j - z_k +
(-\Delta_{j-1} + \Delta_{k+1})/2) \nonumber\\ && - Q(z_j - z_k + (\Delta_j -
\Delta_k)/2) \qquad {\rm for}\ j>k \;,\nonumber\\ L_{jj} &=& 2Q((\Delta_{j-1} +
\Delta_j)/2) \;,
\end{eqnarray}
where $Q(z)$ is given in Eq.~(\ref{qz}). To make further progress, and also to
correct a misprint in Ref.~\cite{egger94}, we now denote the spins
$\sigma_{2q+2-j}$ for $1 \le j \le q+1$ (residing on the backward real-time
branch of $\gamma$) by $\sigma'_j$, and, similarly, the imaginary-time spins
$\sigma_{2q+m}$ for $2 \le m \le r$ by $\bar{\sigma}_m$. In the next step,
employ the coordinate transformation
\[
\eta_j = {1\over2}(\sigma_j + \sigma'_j) \;, \qquad \xi_j = {1\over2}(\sigma_j
- \sigma'_j) \;.
\]
These are ``classical'' and ``quantum'' variables for electronic motion
\cite{egger94}. Exploiting the symmetry relations $Q(z-i\hbar\beta) = Q(-z)$
and $Q(z^\ast) = Q^\ast(-z)$, one finally arrives at
\begin{eqnarray} \label{inflfunct2}
\lefteqn{ \Phi[\{\bar{\sigma}_m,\eta_j,\xi_j\}] = \sum_{m>n=2}^r
\bar{\sigma}_m Y_{mn} \bar{\sigma}_n - {1 \over 4} \sum_{j=1}^q
\sigma_j\sigma'_j \Lambda_{jj} } \nonumber\\
&& + \sum_{j>k=1}^q \xi_j \left( \Lambda_{jk} \xi_k + i X_{jk}
\eta_k \right) + \sum_{j=1}^q\sum_{m=2}^r \xi_j Z_{jm}
\bar{\sigma}_m \nonumber\\
&& + {\sigma'_1 \over 4}
 \sum_{m=2}^r \bar{\sigma}_m (L_{2q+m,1} + L_{2q+m,2q+1}) \;,
\end{eqnarray}
with the matrices
\begin{eqnarray*}
Y_{mn} &=& {1\over4}L_{2q+m,\,2q+n} \,,\qquad
\Lambda_{jk} = {\rm Re}\ L_{jk} \,,\\
X_{jk} &=& {\rm Im}\ L_{jk} \,,\qquad
Z_{jm} = {1\over2}L_{j,2q+m}
\end{eqnarray*}
where $1\le j,k\le q$ and $2\le m,n\le r$. Equation (\ref{inflfunct2}) together
with the contribution from the TLS propagator (\ref{freeprop}) constitutes a
discretized form of the total action which is useful for PIMC
simulations. Finally, to obtain $P(t)$ instead of $C(t)$, the imaginary-time
spin contributions in Eq.~(\ref{inflfunct2}) and in the free propagator have to
be neglected, with $\sigma_1=\sigma'_1=+1$ reflecting the initial
non-equilibrium preparation. Instead of sampling an entirely new MC trajectory
according to this new weight, $P(t)$ can also be calculated from the same MC
trajectory as $C(t)$ by including a correction factor which accommodates the
changes in the corresponding MC weight. We refer the interested reader to the
Appendix for computational details.  Although the latter approach exhibits
poorer statistics, it still produces satisfying results in most cases. We
therefore use it for the simultaneous calculation of $k_f(t)$ and $P(t)$,
without significant increase in computing time.

Unfortunately, this method suffers from the notorious sign problem. It stems
from interference due the complex phase factors assigned to different spin
paths $\{\sigma_j\}$, resulting in a small signal-to-noise ratio of the
stochastic averaging procedure. The exponential increase of interfering paths
with system size $N$ is reflected in an exponential decrease of the
signal-to-noise ratio with the maximum real time under study. Here we employ
the multilevel (MLB) approach in a version suitable to deal with long-ranged
interactions along the (complex) time contour, which is capable of relieving
the sign problem without introducing approximations. The algorithm is described
in detail in Ref.~\cite{mlb00}, and we only give the basic idea in the
following paragraph. For the expert reader, some improvements over
Ref.~\cite{mlb00} are summarized in the Appendix.

In short, the MLB algorithm is based on the simple observation that the sign
problem can be considerably reduced by sampling {\sl blocks} instead of single
paths. A block is here built from a small set of paths. Provided the blocks are
chosen small enough, stochastic averages within a block do not suffer from the
sign problem. Sampling blocks instead of single paths can in fact be shown to
always result in a better signal-to-noise ratio \cite{acp}. Repeating this
scheme iteratively by establishing a hierarchy (``levels'') of blocks, the
exponential severity of the sign problem with respect to the maximum real time
$t$ can be changed into an only algebraic one \cite{mlb98,mlb99,mlb00,mischa}.

Single spin flips and, on the real-time axis, double spin flips, with
simultaneous flips of the forward and backward spins $\sigma_i$ and
$\sigma'_i$, were used to propagate the MC trajectory. Furthermore, since the
creation of short blips (connected real-time intervals where $\xi_i = 0$) is
often energetically more expensive than the creation of more extended ones, we
also included flips of complete blocks of subsequent spins to ensure
ergodicity. When working with MLB, these blocks must not extend over the
borderline separating different levels. However, since the results do not
deviate from those obtained with local spin flips alone, we conclude that
ergodicity poses no problem for our approach.

\section{Numerical results} \label{results}

In this section, we present numerical results for the thermal transfer
rate $k_{th}$ and the occupation probability $P(t)$ of an unbiased system,
$\epsilon = 0$. Of course, then $k_{th}$ is just twice the forward rate
$k_f$. Accordingly, $k_{th}$ was extracted from the plateau value of
\begin{equation}
k(t) = 2k_f(t)\;.
\end{equation}
Furthermore, the right-hand side of Eq.~(\ref{timerate2}) reduces to ${\rm
Im}\{C(t)\}/\hbar\beta$ since $\langle\sigma_z\rangle_\beta$ vanishes for
$\epsilon=0$.  Our code was carefully tested for (a) $\alpha = 0$ (free TLS),
(b) $\alpha = 1/2$ within the scaling limit (\ref{scallimit}), and (c) for
$\alpha = 5,\;\omega_c=0.5\Delta,\; T=2\hbar\Delta/k_B$. The numerical results
accurately reproduce the available analytical solutions [for (a) and (b)] and
the numerical results of Ref.~\cite{thoss} in case (c).  For more tests, see
Sec.~\ref{special}.

We are then interested in mapping out the ET dynamics {\sl phase diagram},
which is a function of three parameters. These are (1) the temperature $T$, (2)
the reorganization energy $\Lambda$, where for the bath spectral density
(\ref{ohmic}) one has $\Lambda=2\alpha \omega_c$ with a damping parameter
$\alpha$ and a cutoff (dominant) frequency $\omega_c$, and (3) the adiabaticity
parameter $\Delta/\omega_c$ with the tunnel splitting $\Delta$. We shall
systematically tune these parameters in order to explore the crossover from
nonadiabatic to adiabatic ET. To keep the relevant parameter space manageable,
we restrict ourselves to the reorganization energy $\Lambda=10\Delta$, but also
give some results for $\Lambda=50\Delta$ \cite{foot}. Unless noted otherwise,
below $\Lambda=10\Delta$ has been taken. We are then left with only two free
parameters, namely temperature and the adiabaticity $\Delta/\omega_c$. By
systematically increasing the latter, one can go from the nonadiabatic into the
adiabatic region of the phase diagram. Of course, we use the term ``phase
diagram'' not in the sense of distinct regions separated by phase transitions.
Nevertheless, there are regimes where ET dynamics proceeds qualitatively
different, which are separated by a rather sharp crossover.

The thermal transfer rate was numerically obtained in two different ways. If
$k(t)$ showed a plateau for some plateau time $t_{\rm pl} > \tau_{\rm trans}$,
the thermal transfer rate $k_{th}$ was taken as the plateau value $k(t_{\rm
pl})$. A second and distinct possibility is to fit $P(t)$ to an exponential
decay $\sim \exp(-k_{th} t)$ after the initial transient, $t>\tau_{\rm
trans}$. In fact, we will encounter examples where a rate from $k(t)$ does
apparently not exist, or at least requires a modified extraction procedure,
while $P(t)$ still displays a well-defined exponential decay on the timescale
under study.

\subsection{Special cases} \label{special}

\begin{figure}
\epsfxsize=6cm
\epsffile{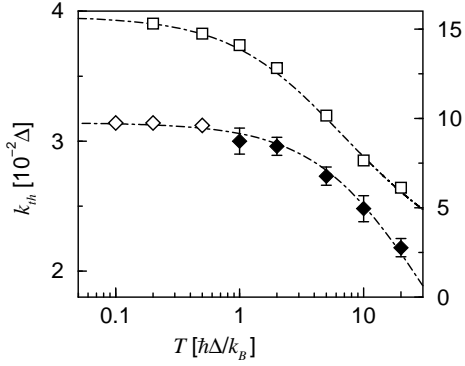}
\caption[]{\label{alpha=0.5_wc=50} Thermal transfer rate $k_{th}$ for
$\alpha=1/2$, $\omega_c=50\Delta$ (diamonds, left axis) and $\omega_c=10\Delta$
(squares, right axis). Filled symbols refer to values obtained from $k(t)$,
open ones to those obtained from $P(t)$. Vertical lines indicate error bars
from statistical sampling uncertainties. The error in fitting $P(t)$ to an
exponential decay is of the order of the symbol size. The dashed-dotted curves
represent the nonadiabatic result (\ref{a12}).}
\end{figure}

To make contact to analytical theory, let us first describe results for the
special damping strengths $\alpha=1/2$ and $\alpha=1$. This provides also
another check for the numerical MLB-PIMC scheme. In addition, from our data for
$\alpha=1/2$, we can give fairly accurate estimates for the validity of the
scaling regime, namely
\begin{equation} \label{scaling}
\omega_c/\Delta \gapx 6 \, \qquad \hbar\omega_c/k_B T \gapx 50\;.
\end{equation}
Let us start with $\alpha=1/2$ and $\omega_c=50\Delta$, corresponding to the
large reorganization energy $\Lambda=50\Delta$. Here an almost strict
exponential decay of $P(t)$ was found at all temperatures, with the rate nicely
matching the analytical prediction (\ref{a12}), see Fig.~\ref{alpha=0.5_wc=50}.
This agreement is of course not surprising since $\omega_c$ is much larger than
$\Delta$. For $P(t)$, transient short-time dynamics reflecting $d{P}/dt(0)=0$
takes place on a significantly shorter timescale than for $k(t)$, where $t_{\rm
pl}$ increases for lower temperatures. For $T\lapx \hbar\Delta/k_B$, $k(t)$
fails to exhibit a plateau while $P(t)$ still allows for a useful rate
description.  This shows the increasing significance of electronic coherence in
this parameter regime \cite{weiss}.

\begin{figure}
\epsfxsize=6cm
\epsffile{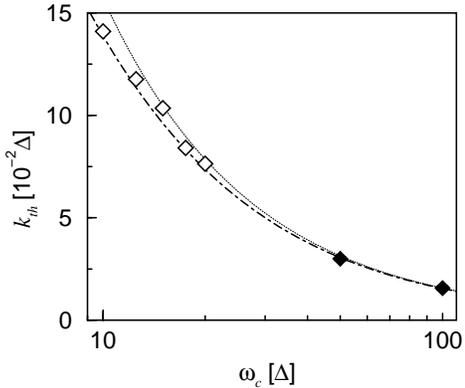}
\caption[]{\label{alpha=0.5_hbeta=1} Same as Fig.~\ref{alpha=0.5_wc=50} but as
a function of $\omega_c/\Delta$ for fixed $T=\hbar\Delta/k_B$.  The dotted
curve denotes the rate according to the scaling limit, while the dashed-dotted
one gives Eq.~(\ref{a12}).}
\end{figure}

This point becomes even more obvious for $\Lambda=10\Delta$, where
$\Delta/\omega_c =0.1$. Now $k(t)$ fails to exhibit a plateau at all
temperatures studied, $0.2 \le k_B T/\hbar \Delta \le 10$.  However, the rate
obtained from $P(t)$ still closely follows the nonadiabatic prediction
(\ref{a12}), see Fig.~\ref{alpha=0.5_wc=50}. In addition, for $\alpha=1/2$ and
$T=\hbar\Delta/k_B$, the dependence on the adiabaticity parameter
 was studied within the range $10 \le \omega_c/\Delta \le
100$. Again the nonadiabatic prediction is nicely reproduced, proving its
usefulness in this regime, see Fig.~\ref{alpha=0.5_hbeta=1}.  The excellent
agreement of Eq.~(\ref{a12}) with our results allows for the quantitative
identification of the scaling regime.  The rate (\ref{a12}) reaches the scaling
value $\pi\Delta^2 / 2\omega_c$ within a relative error of $r$~\%, if
\begin{equation}
{\hbar\omega_c \over k_B T} = {423 \over r} \;.
\end{equation}
Our data indicate that this criterion also holds for the numerically exact ET
rate for (at least) $0.1 \le r \le 10$, $0.01 \le \Delta/\omega_c \le 0.2$, and
$0.05 \le k_B T/\hbar \Delta \le 20$.  Note that this also agrees with
Eq.~(\ref{scaling}) for $r \approx 8.5$~\%.

\begin{figure}
\epsfxsize=6cm
\epsffile{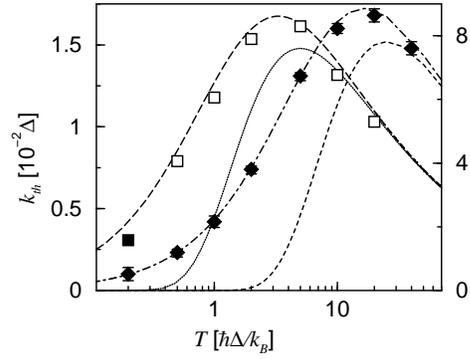}
\caption[]{\label{alpha=1_rate} Thermal transfer rate $k_{th}$ for $\alpha=1$
with $\Lambda=50\Delta$ (diamonds, left axis) and $\Lambda=10\Delta$ (squares,
right axis). The curves give Eq.~(\ref{a1}).  }
\end{figure}

\begin{figure}
\epsfxsize=6cm
\epsffile{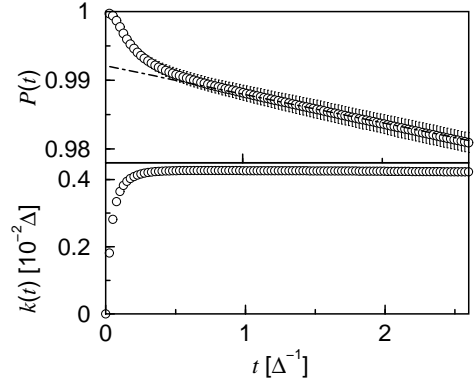}
\hspace{1cm}\vspace{0.5cm}
\caption[]{\label{alpha=1_dynamic} $P(t)$ and $k(t)$ for $\omega_c/ \Delta=25$,
$\Lambda=50\Delta$, and $T=\hbar\Delta/k_B$. The dashed-dotted line represents
$\exp(-0.0042\Delta\,t)-0.008$, in agreement with the plateau value of
$k(t)$. }
\end{figure}

The second special case, to which we now turn, is defined by $\alpha=1$. Again
starting with $\Lambda=50\Delta$, the obtained rate closely follows the
nonadiabatic prediction (\ref{a1}) for all temperatures investigated, $0.2 \le
k_B T/\hbar \Delta \le 20$, see Fig.~\ref{alpha=1_rate}, whereas the classical
(high-temperature) Marcus rate (\ref{marcus}) fails to give an accurate picture
except for $T \ge 40\hbar\Delta/k_B = 1.6 \hbar\omega_c/k_B$. The absence of
electronic coherence and the nice separation of transient and relaxation
timescales is reflected in the fact that $k(t)$ reaches a plateau within the
whole temperature range. The corresponding thermal rate constant is identical
to the rate extracted from the exponential decay of $P(t>\tau_{\rm
trans})$. Moreover, in contrast to $\alpha=1/2$, this exponential decay is
preceded by a steep decrease of the population for $T \le 5\hbar\Delta/k_B$
on timescales $t\lapx \tau_{\rm trans} \simeq 0.5\Delta^{-1}$. The timescale
$\tau_{\rm trans}$ (approximately) coincides for both $P(t)$ and $k(t)$, see
Fig.~\ref{alpha=1_dynamic}, and remains fairly constant with temperature. The
$\alpha=1$ rate for given other parameters is smaller than for
$\alpha=1/2$. Since with our conventions, $\alpha=1/2$ corresponds to a ``more
nonadiabatic'' regime, this finding is not surprising. Recrossing events, which
are irrelevant in the nonadiabatic limit, will generally decrease the rate.

For $\Lambda=10\Delta$, we find a similar scenario, with the Marcus rate
failing except for $T \ge 10\hbar\Delta/k_B = 2\hbar\omega_c/k_B$. However,
$k(t)$ does not exhibit a plateau, since  ET dynamics is now very fast
and hence $\tau_{\rm trans} \ll \tau_{\rm relax}$ does not hold anymore.
Remarkably, $P(t>\tau_{\rm trans})$ still shows exponential decay with the
nonadiabatic rate (\ref{a1}), see Fig.~\ref{alpha=1_rate}.  Note that here the
adiabaticity parameter is already as large as $\Delta/\omega_c=0.2$.

\subsection{Crossover regime}

\begin{figure}
\epsfxsize=6cm
\epsffile{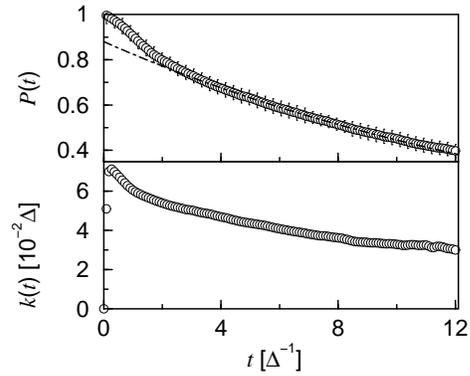}
\hspace{1cm}
\caption[]{\label{alpha=5_dynamic} Same as Fig.~\ref{alpha=1_dynamic} but for
$\Lambda=10\Delta$, $\Delta/\omega_c=1$, and $T=3.333\hbar\Delta/k_B$. The
dashed-dotted line represents $\exp(-0.057\Delta\,t)-0.12$.}
\end{figure}

\begin{figure}
\epsfxsize=6cm
\epsffile{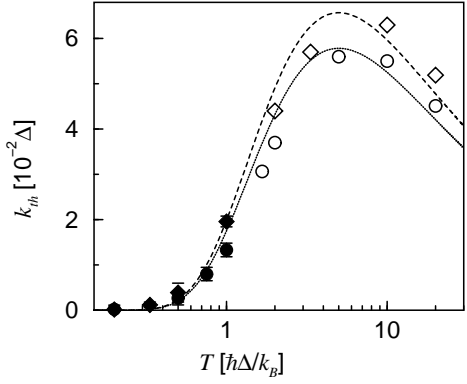}
\caption[]{\label{alpha=5_rate} ET rate $k_{th}$ for $\Lambda=10\Delta$, with
$\omega_c=\Delta$ (diamonds) and $\omega_c=0.5\Delta$ (circles). The dashed and
solid curves represent the corresponding Marcus rates (\ref{marcus}) with
$\omega_r=\omega_c/2$.}
\end{figure}

Entering the crossover regime between nonadiabatic and adiabatic ET (taking
$\Lambda=10\Delta$), let us start with $\Delta/\omega_c=1$. In this case, the
extraction of the thermal transfer rate from $k(t)$ fails for high
temperatures, $T \gapx 2\hbar\Delta/k_B$. After an initial peak, $k(t)$
decreases almost linearly instead of exhibiting a plateau.  The
high-temperature absence of a plateau may be rationalized by noting that the
classical activation barrier (\ref{Fstar}) corresponds to $T =
2.5\hbar\Delta/k_B$. For higher temperatures, ET is practically activationless,
so that transient and relaxation timescale are not really separated anymore.
However, after a rapid initial transient decay, $P(t>\tau_{\rm trans})$ decays
exponentially with the rate $k_{th}$.  Interestingly, this rate coincides with
the value of $k(t)$ at the transition from the initial peak to the linear
decrease, see Fig.~\ref{alpha=5_dynamic}, and governs the behavior of $P(t)$
for $\tau_{\rm trans} \lapx t \lapx \tau_{\rm trans} + 0.5 k_{th}^{-1}$ (note
the increase of the relevant time interval for low temperatures.)  The
temperature dependence of $k_{th}$ is well described by the Marcus formula
(\ref{marcus}), see Fig.~\ref{alpha=5_rate}. While the Marcus rate together
with Eq.~(\ref{omegar}) yields a quite accurate description, the nonadiabatic
rate (\ref{k3}) overestimates the true rate by a factor $\approx 1.2$. However,
overall features (like the temperature at which the maximum of the rate occurs)
are still reproduced. This overestimate of the rate can be rationalized in
terms of recrossing events between acceptor and donor, which are higher-order
contributions in $\Delta$.

\begin{figure}
\epsfxsize=6cm
\epsffile{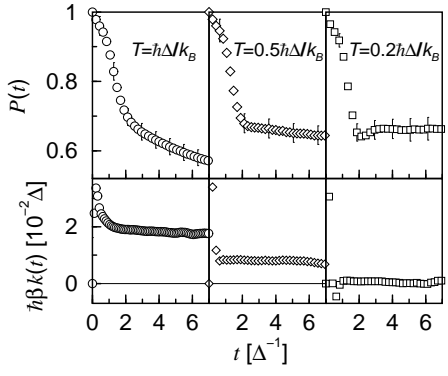}
\hspace{1cm}\vspace{0.5cm}
\caption[]{\label{alpha=5_transient} Same as Fig.~\ref{alpha=5_dynamic} but for
lower temperatures.  Data for $T=0.1\hbar\Delta/k_B$ (not shown) are virtually
identical to those for $T=0.2\hbar\Delta/k_B$, indicating that the
zero-temperature limit has been reached.}
\end{figure}

For low temperatures, $T \lapx 0.2\hbar\Delta/k_B$, we have encountered a
rather interesting and apparently quite general phenomenon regarding the ET
dynamics in the crossover regime. In fact, in this low-temperature crossover
regime between nonadiabatic and adiabatic ET, the thermodynamic rate becomes so
small that almost no change in $P(t>\tau_{\rm trans})$ is observed anymore. At
the same time, the initial decay of $P(t)$ is ended by a local minimum
deepening with decreasing temperature, see Fig.~\ref{alpha=5_transient}.
Therefore the relevant ET dynamics solely happens during the initial
(transient) stage, which lasts for a few $\Delta^{-1}$, while the subsequent
approach to equilibrium takes place on an extremely slow timescale. This {\sl
freezing of the ET dynamics} was never observed in the nonadiabatic regime, but
prevails both in the crossover and the adiabatic regime. Below we refer to this
novel behavior as ``transient ET,'' in contrast to the conventional limiting
cases of adiabatic and nonadiabatic ET.

\begin{figure}
\epsfxsize=6cm
\epsffile{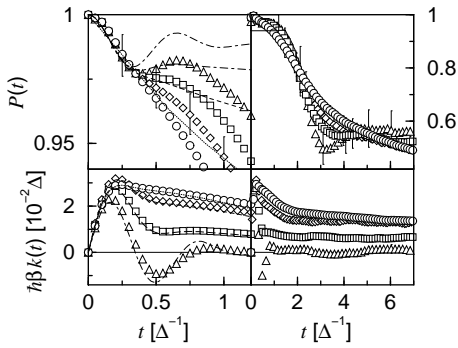}
\hspace{1cm}\vspace{0.2cm}
\caption[]{\label{dble_decay} Same as Fig.~\ref{alpha=5_transient}, but for
$\Delta/\omega_c=2$, $T=2\hbar\Delta/k_B$ (circles), $T=\hbar\Delta/k_B$
(diamonds), $T=0.6\hbar\Delta/k_B$ (squares) and $T=0.2\hbar\Delta/k_B$
(triangles).  The left side of the graph shows the initial transient dynamics.
The box in the top left corner of the upper right graph represents the
corresponding scale of the upper left graph.  The dotted [dashed, long dashed,
dot-dashed] curve represents the adiabatic approximation for
$T=2\hbar\Delta/k_B$ [$\hbar\Delta/k_B$, $0.6\hbar\Delta/k_B$,
$0.2\hbar\Delta/k_B$].}
\end{figure}

Next we increase the value of $\Delta/\omega_c$ to 2. The numerical results in
Fig.~\ref{dble_decay} reveal a similar picture. Again $k(t)$ fails to exhibit a
clear plateau for $T \gapx \hbar\Delta/k_B$, but $k_{th}$ can be extracted from
$P(t)$.  The Marcus formula (\ref{marcus}) captures the temperature dependence
of $k_{th}$ quite well, see Fig.~\ref{alpha=5_rate}, while the nonadiabatic
prediction still reproduces qualitative features, now overestimating the rate
by a factor $\approx 1.4$. Adiabatic dynamics \cite{carmeli} already describes
the transient dynamics, underlining adiabatic tendencies in the bath, but no
oscillations are observed (within error bars). For $T\lapx 0.6\hbar\Delta/k_B$,
$P(t)$ develops a local minimum-maximum pair in the short-time domain $t\lapx
\Delta^{-1}$, whose amplitude increases as $T\to 0$. For $T\lapx
0.3\hbar\Delta/k_B$, we again observe transient ET as described above.
Apparently, ET dynamics becomes quite complex in this regime.

\subsection{Adiabatic limit}

The transition into the adiabatic regime was studied by increasing
$\Delta/\omega_c$ up to 20 for two temperatures, $T = 10\hbar\Delta/k_B$ and
$\hbar\Delta/k_B$. Both $P(t)$ and $k(t)$ are expected to eventually follow the
adiabatic prediction \cite{carmeli} for large $\Delta/\omega_c$, showing damped
oscillations with decreasing amplitude for lower temperatures. These
oscillations are not due to electronic coherence but a consequence of
``nuclear'' (vibrational) coherence. Although adiabatic theory correctly gives
$dP/dt(0)=0$, it is well-known {\sl not}\ to reproduce the correct long-time
dissipative behavior \cite{weiss}. Nevertheless, for $t \ll \tau_{\rm relax}$,
conventional wisdom expects that it should still provide a useful
approximation.  
The short-time dynamics of both $P(t)$ and $k(t)$ indeed closely follows adiabatic theory.
For $T = \hbar\Delta/k_B$, a plateau in $k(t)$ is found, yielding a transfer
rate $k_{th}$ that decreases with increasing $\Delta/\omega_c$, until
oscillations eventually occur at $\Delta/\omega_c \gapx 5$.  
Since (within error bars) $P(t)$ shows no oscillations even for
$\Delta/\omega_c=5$, we can assign a rate to ET under a nonequilibrium initial
preparation.  Only for larger $\Delta/\omega_c$, evidence for oscillations in
$P(t)$ can be observed, see also Ref.~\cite{mlb00}.

\begin{figure}
\epsfxsize=6cm
\epsffile{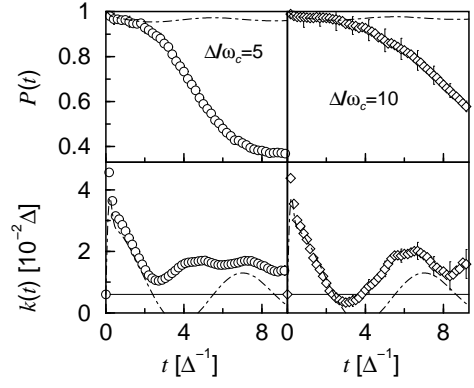}
\caption[]{\label{adiabatic_dynamics_low_T} $P(t)$ and $k(t)$ for
$T=\hbar\Delta/k_B$, $\Delta/\omega_c=5$ (left) and $\Delta/\omega_c= 10$
(right).  Dot-dashed curves represent the adiabatic prediction. }
\end{figure}

\begin{figure}
\epsfxsize=6cm
\epsffile{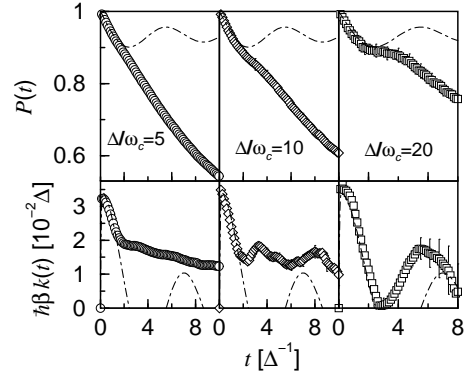}
\caption[]{\label{adiabatic_dynamics_high_T} Same as
 Fig.~\ref{adiabatic_dynamics_low_T}, but for $T=10\hbar\Delta/k_B$ and (from
 left to right) $\Delta/\omega_c=5$, 10 and 20.  }
\end{figure}

Several important differences to adiabatic theory appear in our data, see
Figs.~\ref{adiabatic_dynamics_low_T} and \ref{adiabatic_dynamics_high_T}.  With
regard to $k(t)$, the observed oscillations have a smaller amplitude, their
mean value is larger, and the oscillation frequency is higher.  Adiabatic
theory assumes a static bath, which under a dynamical
description implies incorrect long-time behavior and overly pronounced
oscillations.  This qualitatively explains the first two differences.  The
increase of the oscillation frequency is then due to the larger energy
difference between the two adiabatic potential surfaces away from the barrier
region, since the region away from the barrier top is probed extensively by a
dynamical bath. Turning next to $P(t)$, adiabatic theory fails to
reproduce the observed steep descend at the end of the transient dynamics. This
can be explained in terms of the bath distribution. In adiabatic theory, the
distribution remains centered around the donor state, see
Eq.~(\ref{initial_perp_P}), so that there is a low probability of reaching the
transition region, and hence too slow ET dynamics is predicted.  
The improved adiabatic description \cite{carmeli} captures the damping
of the oscillation amplitude and the increase of the frequency much more accurately.
However, since it is still based on an average over a static bath distribution,
it again misses the steep descend of $P(t)$ and the correct mean value
for $k(t)$ at times $t>t_{\rm trans}$.
For increasing $\Delta/\omega_c$, adiabatic theory then provides a
systematically better description,
although substantial differences remain even at $\Delta/\omega_c=20$.  For $T =
10\hbar\Delta/k_B$, we get a similar picture, see
Fig.~\ref{adiabatic_dynamics_low_T}.  However, as expected from our above
discussion, $k(t)$ does not exhibit a plateau.  The transition from incoherent
to coherent (oscillatory) behavior in $P(t)$ takes place between
$\Delta/\omega_c = 5$ and 10.  Remarkably, true adiabatic dynamics is realized
only for about half an oscillation period, this being virtually independent of
$\Delta/\omega_c$. This shows that the long-time relaxation process not
captured by adiabatic theory sets in much earlier than thought previously.

\subsection{Validity of rate description}

\begin{figure}
\epsfxsize=6cm
\epsffile{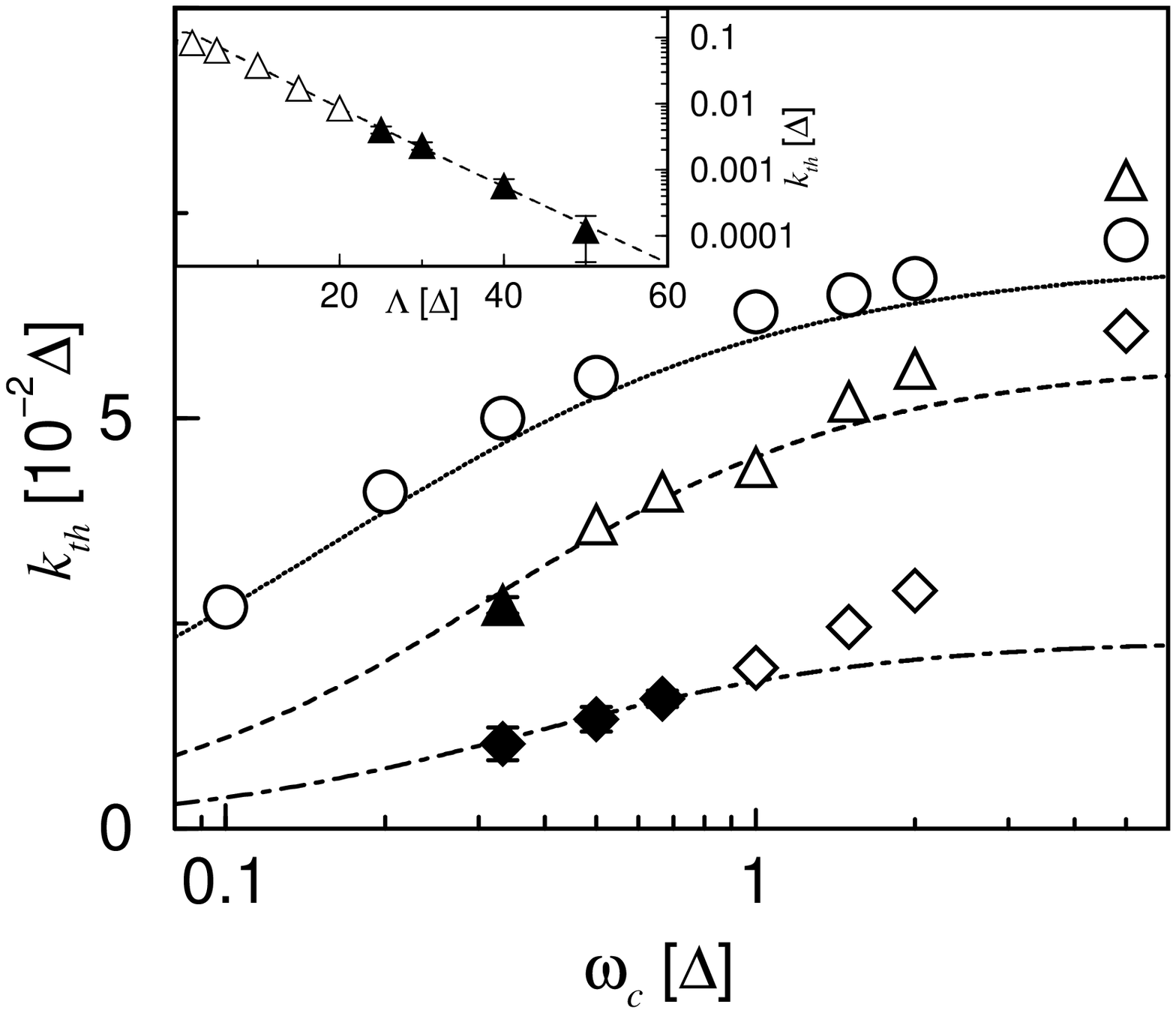}
\caption[]{\label{omega_r} Thermal transfer rate $k_{th}$ as a function of
$\omega_c$ for $T=10\hbar\Delta/k_B$ (circles), $T=2\hbar\Delta/k_B$
(triangles) and $T=\hbar\Delta/k_B$ (diamonds). The dotted [dashed,
dashed-dotted] lines refer to the Marcus rate with $\omega_r$ according to
Eq.~(\ref{wr}) and $q(T=10\hbar\Delta/k_B)=0.0002$
[$q(T=2\hbar\Delta/k_B)=0.21$, $q(T=\hbar\Delta/k_B)=0.26$]. The exact rates
exceed the nonadiabatic (maximum) Marcus rate for $\omega_c \gapx 5\Delta$
[$2\Delta$, $1.5\Delta$]. The inset shows $k_{th}$ at $T=2\hbar\Delta/k_B$ and
$\omega_c = 0.5\Delta$ as a function of the reorganization energy $\Lambda$,
where the dashed curve corresponds to the main graph.}
\end{figure}

Having identified the threshold to the adiabatic regime (at least for
$\Lambda=10\Delta$ and $k_B T/\hbar\Delta = 1, 10$), we next comment on the
validity of a rate description for ET.  A rate description for $P(t)$ turns out
to be appropriate for $\Delta/\omega_c \lapx 5$, where accurate fits of the
numerically exact rates to analytical theory are possible. For small
$\Delta/\omega_c$, the rate is given by the nonadiabatic prediction (\ref{k3}),
while in the crossover regime, the Marcus rate (\ref{marcus}) is superior. The
Marcus formula assumes a classical bath, and hence becomes more appropriate for
large $\Delta/\omega_c$, where the bath is very slow and hence classical.
Remarkably, the Marcus formula can be made to work even outside the true
classical (high-temperature) regime, see Fig.~\ref{omega_r}. In the
high-temperature limit, the solvent frequency $\omega_r$ appearing in
Eq.~(\ref{marcus}) is given by Eq.~(\ref{omegar}). For lower (but not too low)
temperatures, the Marcus formula still provides a good estimate for the exact
rate within the regime stated above, provided $\omega_r$ is taken in the form
\begin{equation}\label{wr}
\omega_r = {\omega_c \over 2} \left({\omega_c \over \Lambda}\right)^{q(T)} \;.
\end{equation}
In the high-temperature limit, our data indicate that Eq.~(\ref{wr}) reduces to
Eq.~(\ref{omegar}), implying $q(T) \to 0$ for high temperatures.  At low
temperatures, however, a positive value for $q(T)$ is found, see
Fig.~\ref{omega_r}.  Of course, at sufficiently low temperatures, the
quantum-mechanical ``transient regime'' mentioned above is entered, where the
Marcus formula breaks down.  Nevertheless, Eq.~(\ref{wr}) represents a
noticeable improvement over Eq.~(\ref{omegar}) for a significant range of
intermediate temperatures.  We term this low-temperature region, where
classical Marcus theory along with Eq.~(\ref{wr}) becomes accurate, ``extended
Marcus regime.'' That such a regime could indeed exist may be rationalized by
noting that for $\Delta/\omega_c\gapx 1$, the bath behaves already rather
classically, yet quantum fluctuations significantly renormalize the
solvent frequency scale $\omega_r$ at low-to-intermediate temperatures.  It would
clearly be interesting to provide an analytical derivation for Eq.~(\ref{wr}).

Notably, for $\Delta/\omega_c\gapx 10$, no rate description is possible
anymore, neither based on $P(t)$ nor on $k(t)$, since ET dynamics is dominated
by vibrational coherence. The rate formalism also becomes useless at very low
temperatures, $T \lapx 0.2\hbar\Delta/k_B$, within the crossover regime. As
elaborated above, in this ``transient regime,'' $k_{th}$ is almost zero such
that long-time relaxational ET is essentially frozen in. The important ET
dynamics then happens during the initial transient, $t \lapx \tau_{\rm
trans}$. Here our simulations for $P(t)$ reveal a quite complex behavior.
After a very slow initial onset, a fast transient is found, which eventually
ends in a local minimum marking the transition to the frozen relaxation regime,
$k_{th} \approx 0$.  This transient regime does not extend into the
nonadiabatic region. For instance, for $T=0.2\hbar\Delta/k_B$, it could only be
observed for $\Delta/\omega_c \gapx 0.2$.

\section{Conclusions and Discussion} \label{conclusion}

\begin{figure}
\epsfxsize=6cm
\epsffile{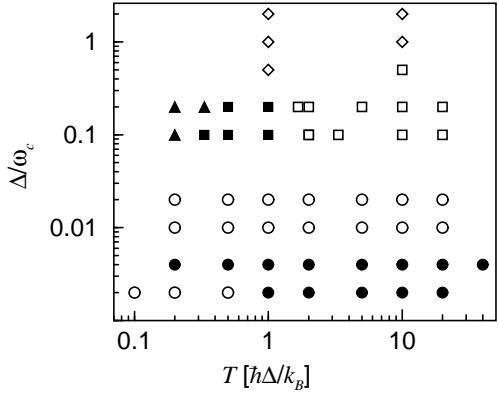}
\hspace{2.5cm}
\caption[]{\label{parameterspace} Location of parameter values where
simulations  have been carried out.  The reorganization energies are
 $\Lambda=50\Delta$
for $\Delta/\omega_c=0.02, 0.04$, and $\Lambda=10\Delta$ otherwise.
Results following the nonadiabatic prediction (\ref{k3}) are denoted by
circles, those following the (classical or extended) Marcus formula (\ref{marcus}) by squares.
Transient behavior is marked by triangles, and oscillatory (adiabatic)
dynamics by diamonds.
}
\end{figure}

Using the spin-boson model as a description for ET processes, we have
calculated the thermal transfer rate $k_{th}$ for two different reorganization
energies, $\Lambda=10\Delta$ and $\Lambda=50\Delta$, where $\Delta$ is twice
the electronic coupling between the two redox states. Our real-time PIMC
simulations are able to cover the full range of temperatures and adiabaticity
parameters $\Delta/\omega_c$, where $\omega_c$ denotes a typical prominent bath
frequency. In this paper, we have confined ourselves to symmetric (unbiased) ET
systems.  The location of the calculated rates in parameter space are
schematically shown in Fig.~\ref{parameterspace}.  The time-dependent function
$k(t)$, see Eq.~(\ref{timerate}), is a powerful means both to obtain $k_{th}$
and to decide whether a rate description is appropriate in the first place. In
the absence of electronic or vibrational coherence, this rate also describes
the exponential decay of the electronic population $P(t)$, provided the
timescales for transient dynamics $\tau_{\rm trans}$ and thermal relaxation
$\tau_{\rm relax}$ are sufficiently well separated. However, even in the
absence of such a separation, $P(t)$ can often be described by a rate with good
accuracy. In this case we found that $k_{th}$ can be extracted from $k(t)$ if
the transition between transient motion and the relaxation process can be
clearly identified, this usually being the case if transient motion displays a
pronounced peak.  That there exist differences in the ET dynamics obtained from
$P(t)$ or $k(t)$ should not come as a surprise, since these quantities
correspond to different initial preparations.

In the nonadiabatic regime, Eq.~(\ref{k3}) captures the thermal transfer rate
accurately for $\Delta/\omega_c\lapx 0.1$. Even for $\Delta/\omega_c = 1$ or
$2$, nonadiabatic theory typically overestimates $k_{th}$ by only a small
factor of the order $1$ to $2$, reflecting the increasing relevance of
recrossing events neglected in Eq.~(\ref{k3}). Based on our simulations, we
were also able to give precise bounds for the validity of the so-called scaling
picture, see Eq.~(\ref{scaling}). The crossover to the adiabatic regime is then
surprisingly well described by the famous Marcus theory, see
Eq.~(\ref{marcus}). For high temperatures (within the classical regime), the
solvent scale $\omega_r$ in Eq.~(\ref{marcus}) is given by Eq.~(\ref{omegar}).
For lower temperatures, we identified an ``extended'' Marcus regime, where
$\omega_r$ is empirically given by Eq.~(\ref{wr}).  Eventually, for very low
temperatures, $T\lapx \hbar\Delta/k_B$, ET dynamics is almost completely
restricted to transient motion, which cannot be explained by Marcus theory but
requires a full dynamical description.  In this regime, analytical progress is
most difficult, and our exact data reveal the presence of a ``transient
regime,'' where ET proceeds solely via this fast initial transient, while
long-time relaxation is extremely slow.  Finally, we have shown that for
$\Delta/\omega_c\gapx 10$, ET dynamics is fully coherent, and no rate
description is possible. A comparison to the adiabatic description
\cite{carmeli} reveals that the latter can only explain the dynamics
of both $P(t)$ and $k(t)$ for times up to $\tau_{\rm trans}$.

\begin{figure}
\epsfxsize=6cm
\epsffile{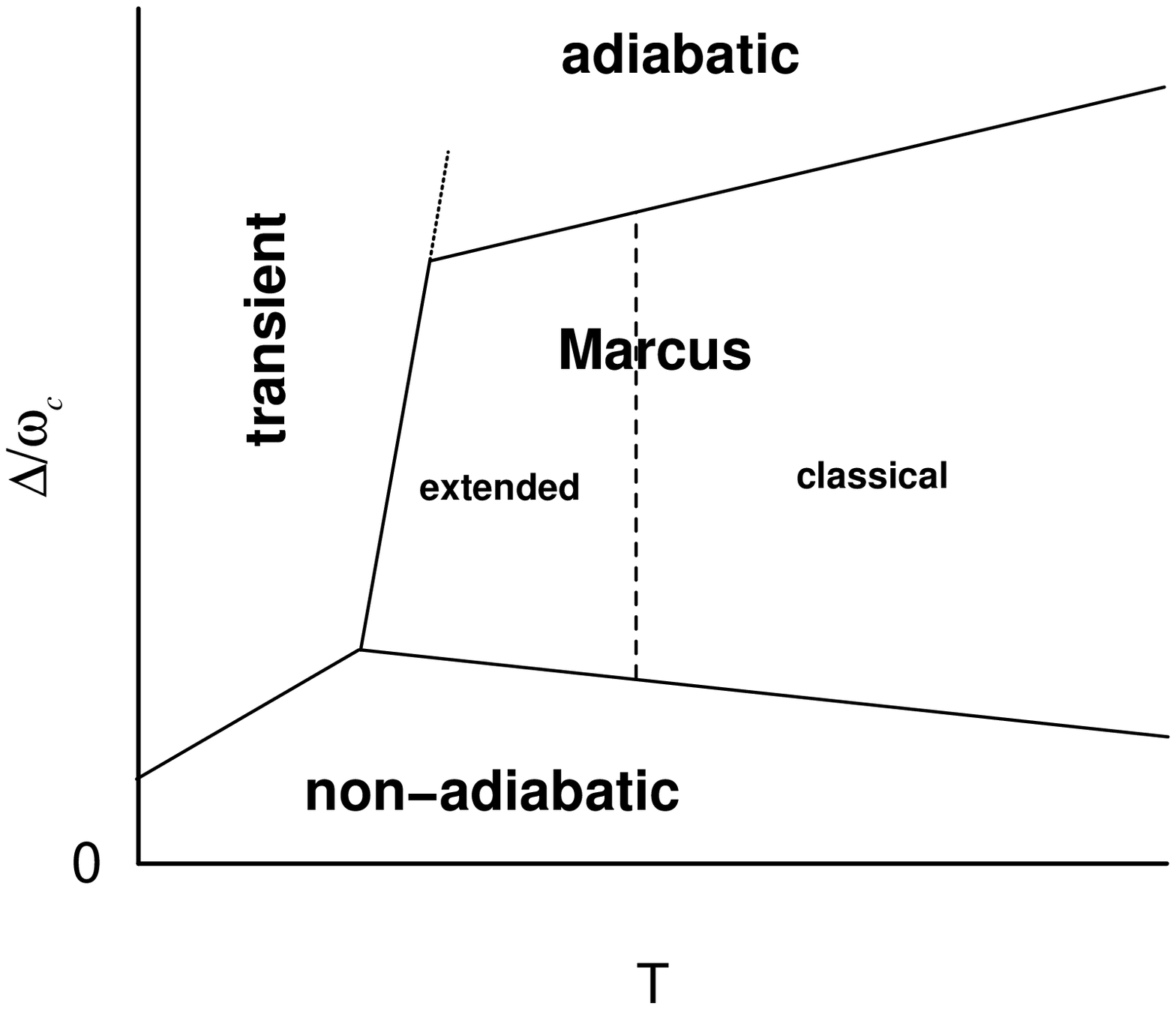}
\hspace{2cm}
\caption[]{\label{phasediag} Schematic ET dynamics ``phase diagram.'' To obtain
quantitative numbers for the crossover lines, refer to
Fig.~\ref{parameterspace}.}
\end{figure}

Our findings are summarized in the schematic ``phase diagram'' depicted in
Fig.~\ref{phasediag}. Within large areas of parameter space, ET can be
described by an exponential decay after a transient timescale, $t > \tau_{\rm
trans}$. Throughout this rate regime, the corresponding rate constant is very
well captured by analytical approaches, namely the nonadiabatic approximation
for small $\Delta/\omega_c$, and Marcus theory for intermediate-to-large
$\Delta/\omega_c$. However, a breakdown of the rate description can be observed
in two different cases. In the strongly adiabatic regime, nuclear (vibrational)
coherence causes oscillatory behavior. In addition, at very low temperatures
within the crossover regime, relaxation is frozen in, and the transient motion
dominates the ET process (``transient regime'').

Clearly, the computational method used in this paper has the potential to solve
many other interesting problems via real-time PIMC. In the context of ET
dynamics, we have not reported results for the biased system, but plan to do so
in the future.

\acknowledgements

We wish to thank J. Ankerhold, M. Dikovsky, H. Grabert, A. Lucke, C.H. Mak, and
J.T. Stockburger for valuable discussions. Financial support by the
Volkswagen-Stiftung (No.~I/73 259) is acknowledged.

\appendix
\section{Computational details}
\label{append}

In this Appendix, some details and improvements over the approach of
Ref.~\cite{mlb00} are summarized. Paths are blocked together in the following
way. The discretized contour $\gamma$ is divided into $L$ consecutive parts
$\{z_1, \dots, z_{q_1}\},\dots,\{z_{q_{L-1}+1}, \dots, z_N\}$, corresponding to
the different levels $l=1,\dots,L$, dividing the spin paths $\{\sigma_j\}$ into
$L$ respective subsets $\{\sigma_j\}_l$. Starting on the first level, for each
configuration of the $\{\sigma_j\}_{l>1}$ the respective level-1 block is
formed by all possible configurations of the sub-path
$\{\sigma_j\}_1$. Switching to the second level, changing only the spins
$\{\sigma_j\}_2$ with all others fixed gives a level-2 block for each
configuration of the $\{\sigma_j\}_{l>2}$. Clearly, each level-2 block contains
a whole set of level-1 blocks. This scheme is continued until on the highest
level $L$, the final sampling of expectation values is carried out, with
strongly reduced sign problem.

To calculate the appropriate block averages, the total action must factorize
with respect to the different levels. An appropriate partitioning of
$\Phi[\{\bar{\sigma}_m,\eta_j,\xi_j\}]$ is given by
\begin{eqnarray}
\lefteqn{\Phi_1[\{\bar{\sigma}_m,\eta_j,\xi_j\}] = \sum_{m=2}^r
\bar{\sigma}_m \left[
 \sum_{n>m}^r Y_{mn} \bar{\sigma}_n \right.} \nonumber\\
 &&\left. + \sum_{j=1}^q \xi_j Z_{jm} +
 {1 \over 4}\sigma'_1 (L_{2q+m,1} + L_{2q+m,2q+1}) \right]
 \nonumber\\
&& + \sum_{k=2}^{q_1}\sum_{j>k}^q \xi_j \left( \Lambda_{jk}
\xi_k + i X_{jk} \eta_k \right) - {\Lambda_{11}\over4}
 \sigma_1\sigma'_1 \nonumber\\
&&- {\Lambda_{22}\over4} \sum_{j=1}^{q_1}
 \sigma_j\sigma'_j \;, \nonumber
\end{eqnarray}
\begin{eqnarray}
\Phi_{l>1}[\{\eta_j,\xi_j\}_{l' \ge l}] &=&
 \sum_{k=q_{l-1}+1}^{q_l}\sum_{j>k}^q \xi_j \left( \Lambda_{jk} \xi_k + iX_{jk}
\eta_k \right) \nonumber\\
&& - {\Lambda_{22} \over 4} \sum_{j=q_{l-1}+1}^{q_l}
\sigma_j\sigma'_j
\end{eqnarray}
(note $\Lambda_{jj} = \Lambda_{22}$ for $2 \le j \le q$) with
$\Phi[\{\bar{\sigma}_m,\eta_j,\xi_j\}] = \sum_{l=1}^L
\Phi_l[\{\bar{\sigma}_m,\eta_j,\xi_j\}_l]$. The contribution from the free
propagator is split in the same manner. Here we include the imaginary-time
spins into the first level, since they contribute only
weakly to the overall sign problem.  Therefore they can be assigned to any level of
real-time spins without significantly affecting the sign problem. Since the
numerical effort to update spins increases for high levels, it saves computing
time to put them into the lowest one.

For a computation of $C(t)$, a slight modification of previous MLB formulations
is necessary since in Refs.~\cite{mlb98,mlb99,mlb00,mischa} averages can be
computed only on time slices belonging to the topmost level. This extension of
MLB to a computation of an arbitrary $n$-point function
\begin{equation} \label{A}
A(t_1,\dots,t_n) = \langle A[\{\sigma_j\}]\rangle \simeq Z^{-1}
\sum_{\{\sigma_j\}} A[\{\sigma_j\}] \rho[\{\sigma_j\}]
\end{equation}
is described next. At first a dependence of $A(t_1,\dots,t_n)$ on all time
slices seems to contradict the philosophy underlying MLB, namely tackling the
sign problem by replacing the sampling process over all possible system paths
by one over the level-$L$ block only. However, decomposing $A$ in the same way
as $\Phi$,
\begin{equation}
A[\{\sigma_j\}] = \prod_{l=1}^L A_l[\{\sigma_j\}_{l' \ge l}] \;,
\end{equation}
block averages of $A_l$ for $l<L$ remove all dependence on lower-level spins.
Therefore the remaining accumulation process can again be performed by sampling
on the highest level only. These block averages are calculated as modified
``bonds'' $\hat{B}_l[\{\sigma_j\}_{l'>l}] = A_{l+1}[\{\sigma_j\}_{l'>l}]
B_l[\{\sigma_j\}_{l'>l}]$ where $B_l$ denotes the  ``bond'' of
Ref.~\cite{mlb00},
\begin{equation}
B_l[\{\sigma_j\}_{l'>l}] = C_{l-1}^{-1} \sum_{\{\sigma_j\}_l}
B_{l-1}[\{\sigma_j\}_{l'\ge l}] e^{-W_l(\{\sigma_j\}_{l' \ge l})}
\end{equation}
with the normalization $C_{l-1} = \sum_{\{\sigma_j\}_l} {\cal
P}_{l-1}[\{\sigma_j\}_l]$ of the weight for level-$l$ bonds,
\begin{eqnarray}
{\cal P}_{l-1}[\{\sigma_j\}_l] &=&
\Big|B_{l-1}[\{\sigma_j\}_l,\{\sigma_j\}^0_{l+1},\dots,\{\sigma_j\}^0_L]
\nonumber\\
&& \times
e^{-W_l(\{\sigma_j\}_l,\{\sigma_j\}^0_{l+1},\dots,\{\sigma_j\}^0_L)}\Big|\;.
\end{eqnarray}
Sums run over all possible configuration of level-$l$ spins $\{\sigma_j\}_l$,
and $\{\sigma_j^0\}_{l'<l}$ denotes their initial configuration at the start of
the MC trajectory. Hence $A(t_1,\dots,t_n)$ can be written as
\begin{eqnarray}
A(t_1,\dots,t_n) &=& \frac{\sum_{\{\sigma_j\}_L}
\hat{B}_{L-1}[\{\sigma_j\}^{(i)}_L] e^{-W_L[\{\sigma_j\}^{(i)}_L]}}
{\sum_{\{\sigma_j\}_L} B_{L-1}[\{\sigma_j\}^{(i)}_L]
e^{-W_L[\{\sigma_j\}^{(i)}_L]}} \nonumber\\
&\simeq& \frac{\sum_{i=1}^K {\hat{B}_{L-1}[\{\sigma_j\}^{(i)}_L]
\exp\{-W_L[\{\sigma_j\}^{(i)}_L]\} \over \left|B_{L-1}[\{\sigma_j\}^{(i)}_L]
\exp\{-W_L[\{\sigma_j\}^{(i)}_L]\}\right|}}{\sum_{i=1}^K
{B_{L-1}[\{\sigma_j\}^{(i)}_L] \exp\{-W_L[\{\sigma_j\}^{(i)}_L]\} \over
\left|B_{L-1}[\{\sigma_j\}^{(i)}_L] \exp\{-W_L[\{\sigma_j\}^{(i)}_L)\}\right|}}
\;,
\end{eqnarray}
where the sums in the last expression run over $K$ configurations
$\{\sigma_j\}^{(i)}_L$ of the level-$L$ spins distributed according to ${\cal
P}_L[\{\sigma_j\}_L]$. Here $K$ is the usual MLB sampling number \cite{mlb98},
with typical values $K\approx 200$ employed in the simulations. Therefore
$A(t_1,\dots,t_n)$ is accessible to a MLB scheme which not only soothes the
sign problem due to interfering phase factors of different spin paths, but also
takes care of alternating signs of the lower-level contributions to $A$.

Finally we outline a scheme suitable for compensating for different initial
preparations, i.e. computing $P(t)$ with the same MC trajectories as
$C(t)$. After discretizing the corresponding path integrals, $P(t)$ can be
written as, cp.~Eq.~(\ref{corr3}),
\begin{eqnarray} \label{P2}
P(t) &=& Z_P^{-1} \sum_{\{\sigma_j\}'} \sigma_{q+1} \rho_P[\{\sigma_j\}'] \,,
\nonumber \\
\rho_P[\{\sigma_j\}'] &=& K(+1,\sigma_2)K^\ast(+1,\sigma'_2)\left[
\prod_{i=2}^N K(\sigma_i,\sigma_{i+1}) \right] \nonumber\\
&&\times e^{-\Phi_P[\{\sigma_j\}'] + \zeta[\{\sigma_j\}']} \,,
\end{eqnarray}
where the summation now only includes real-time spins $\sigma(t>0)$,
i.e. $\{\sigma_j\}'=\{\sigma_2,\dots,\sigma_{2q}\}$. The corresponding
influence functional $\Phi_P[\{\sigma_j\}']$ is obtained from
Eq.~(\ref{inflfunct2}) by dropping all terms with imaginary-time spins
$\bar{\sigma}_m$ and setting the $t=0$ spins $\sigma_1, \sigma'_1$ to
$+1$.  In addition
\begin{equation}
\zeta[\{\sigma_j\}'] = i\sum_{j=2}^q \xi_j {\rm Im}\{Q((j-1/2)\Delta_j)-Q((j-3/2)\Delta_j)\}
\end{equation}
includes a contribution describing the initial preparation of the bath
\cite{weiss,mlb00}. To put $P(t)$ into the form of Eq.~(\ref{A}), we
rewrite Eq.~(\ref{corr3}) according to
\begin{equation} \label{P3}
P(t) = 2^{-m-2} Z_P^{-1} \sum_{\{\sigma_j\}} \left( \sigma_{q+1}
{\rho_P[\{\sigma_j\}'] \over \rho[\{\sigma_j\}]} \right) \rho[\{\sigma_j\}]\;,
\end{equation}
with [note that $\sum_{\{\sigma_1,\sigma'_1\}} \sum_{\{\bar{\sigma}_m\}} 1 = 2^2 \cdot
2^m$]
\begin{eqnarray}
2^{-m-2} Z_P^{-1} &=& \left( 2^{m+2}
\sum_{\{\sigma_j\}'} \rho_P[\{\sigma_j\}'] \right)^{-1} \nonumber\\
&=& \left( \sum_{\{\sigma_j\}} {\rho_P[\{\sigma_j\}'] \over
\rho[\{\sigma_j\}]} \rho[\{\sigma_j\}] \right)^{-1} \,.
\end{eqnarray}
The effect of the different initial preparation is now absorbed by the factor
$\rho_P/\rho$. Consequently $P(t)$ can now be computed as described above with
$A[\{\sigma_j\}] = \sigma_{q+1} \rho_P[\{\sigma_j\}']/\rho[\{\sigma_j\}]$. The
prefactor can be treated on the same footing. Thus, with the above extension in
mind, $P(t)$ can be calculated from the same MC trajectories used for
sampling $C(t)$ within our MLB-PIMC approach.

\clearpage

%##############################################################################
% Figures
%##############################################################################

\begin{references}
\bibitem{foot}
We consider only $\alpha \ge 1/2$. ET dynamics is normally incoherent with
respect to electronic motion, while for $\alpha < 1/2$, spin-boson dynamics
shows electronic coherence \cite{weiss}.
\end{references}
\end{document}